\documentclass[prd,superscriptaddress,showpacs,twocolumn,amsmath,amssymb]{revtex4}
\usepackage{graphicx}
\usepackage{amsmath}
\usepackage{amssymb}
\usepackage{float}
\usepackage{color}
 \usepackage{braket}

 \usepackage{appendix}
\usepackage{cancel,xcolor}
\usepackage{indentfirst}
\usepackage{CJKulem}
\usepackage{soul,xcolor}
\setstcolor{red}
\usepackage{ulem}
\usepackage{cancel}
\usepackage[pdfpagemode=UseNone,pdfstartview=FitH,colorlinks=true,linkcolor=blue,urlcolor=blue,anchorcolor=blue,citecolor=blue]{hyperref}

\newcommand\bea{\begin{eqnarray}}
\newcommand\eea{\end{eqnarray}}
\newcommand\beq{\begin{equation}}  
\newcommand\eeq{\end{equation}}

\begin{document}

\title{\titlename}
\title{Krylov state complexity for BMN matrix model}
\author{Dibakar Roychowdhury}
\email{dibakar.roychowdhury@ph.iitr.ac.in}
\affiliation{Department of Physics, Indian Institute of Technology Roorkee
Roorkee 247667, Uttarakhand,
India}

\begin{abstract}
We explore Krylov complexity in the BMN matrix model following a systematic reduction of it, known as the pulsating fuzzy sphere model. We present an analytical setup that allows us to calculate Lanczos coefficients in both large and small deformation limits of the matrix model. 
\end{abstract}
\maketitle
\section{Introduction and motivation}
In recent years, Krylov complexity \cite{Parker:2018yvk}-\cite{Caputa:2024vrn} has drawn renewed attention in the context of quantum chaos \cite{Hashimoto:2023swv}-\cite{Bhattacharjee:2024yxj}, holography \cite{Caputa:2024sux}-\cite{Erdmenger:2022lov} and black holes \cite{Kar:2021nbm}. The basic idea behind Krylov complexity rests on the fact that it minimizes the spread of the wave function under Hamiltonian evolution. In other words, the Krylov basis is a subspace of the Hilbert space where the cost function is minimum. The Hamiltonian is tri-diagonal and the non-zero elements are called the Lanczos coefficients \cite{Balasubramanian:2022tpr}.

The purpose of this paper is to explore the Krylov state complexity in the BMN matrix model \cite{Berenstein:2002jq}, which goes under the name of pulsating fuzzy sphere \cite{Asano:2015eha}-\cite{Huh:2024ytz}. In a complementary paper \cite{Huh:2024ytz}, the above issue has been addressed based on numerics, revealing an integrable structure both in the small and large values of the mass parameter ($\mu$). However, for intermediate $\mu$, the matrix model exhibits a chaotic behavior that is reflected in the plateau followed by a ramp in the Krylov complexity.   

In the present paper, we pursue this problem based on \textit{analytical} techniques. We show that it is possible to set up a calculation in the extreme corners of the (mass) parameter space, that is, for large and small values of $\mu$. In the regime of large $\mu$, the wavefunction looks like a product of two harmonic oscillator wave functions in their ground states that are highly localized around their respective vacuum \cite{Amore:2024ihm}. On the other hand, for $\mu \sim 0$, the matrix model should be interpreted as the BFSS matrix model \cite{Banks:1996vh} with massive perturbations \cite{Dasgupta:2002hx}. 

In the regime of large mass deformation ($\mu \gg 1$), the Lanczos coefficients reveal a universal scaling, namely $a_n \sim \mu$ and $b_n \sim \mu$. The Krylov complexity growth is purely fixed by the mass parameter $\mu$, which can be schematically expressed as an expansion of the form
\begin{align}
    C(t)-C_0 \sim \beta \mu^2 t^2+\gamma \cos (\mu t)+\zeta \mu t \sin(\mu t)+\cdots
\end{align}
where $C_0, \beta, \gamma, \zeta$ are constants and can be expressed in terms of Lanczos coefficients ($a_n,b_n$). The above analysis is performed in detail in Section \ref{sec2}.

Next, we move on to the other corner of the parameter space, where we treat the mass parameter small $\mu \sim 0$ and in particular as a perturbation of the BFSS matrix model \cite{Banks:1996vh}. The wavefunction at $t=0$ corresponds to a free particle, where the initial state corresponds to the minima of the potential. We calculate the Hamiltonian evolution of this initial state (at $t=0$) under the BMN Hamiltonian and calculate the Krylov complexity in the small deformation limit. This analysis is performed in Section \ref{sec3}, where the BMN Hamiltonian is treated as a small $\mu$ expansion about the free particle Hamiltonian. Unlike the strong deformation limit, the LO corrections to the Lanczos coefficients appear in order $\mu^2$. Finally, we draw our conclusion in Section \ref{sec4}.

\section{Large deformation limit}
\label{sec2}
In this Section, we perform a calculation in the limit of large mass deformation, that is, $\mu \gg 1$. Following the scaling arguments of \cite{Amore:2024ihm}, this would correspond to localized harmonic oscillator states around the respective minima $x ,y\sim 0$ of the potential, such that $\mu x$ and $\mu y$ are finite. The subsequent time evolution of this initial state under the BMN Hamiltonian in the Krylov basis is what we identify as the Krylov complexity at large $\mu$.

The $N=2$ matrix model (also known as the pulsating fuzzy sphere model) is characterized by the following Hamiltonian \cite{Asano:2015eha}-\cite{Huh:2024ytz}
\begin{align}
\label{e1}
    &H(x,y)=\frac{p^2_x}{2}+\frac{p^2_y}{2}+V(x,y)\\
    & V (x,y)=\frac{\mu^2}{8}(x^2+ 4 y^2)- \mu y^3+\frac{1}{2}(x^2+y^2)^2
    \label{e2}
\end{align}
where in the limit $\mu \rightarrow 0$, the Hamiltonian \eqref{e1} corresponds to the BFSS matrix model \cite{Banks:1996vh}. 

To begin with, we propose the state at $t=0$ as follows
\begin{align}
\label{e3}
    \psi_0 (x,y)=\sqrt{\frac{2 \mu}{\pi}}e^{-\mu (x^2+y^2)}
\end{align}
where we choose to work in natural units $\hbar =c=1$. Given the potential function \eqref{e2}, we assume that both oscillators have identical mass $\mu$ and different fundamental frequencies $\omega =1$ and $\omega =\frac{1}{2}$.

Notice that the above ansatz \eqref{e3} is valid only for $\mu \neq 0$. The above choice \eqref{e3} comes from the fact that, to begin with, we have two \textit{localized} harmonic oscillators which thereby start to interact at later times ($t>0$). In other words, the harmonic oscillator ansatz \eqref{e3} is a good approximation at low energies when the system stays very close to the minimum of the potential \eqref{e2}. From the point of view of the scaling arguments \cite{Amore:2024ihm}, the above choice \eqref{e3} corresponds to setting $\mu$ large, which also sets the energy of the configuration in its ground state. 

This can be checked explicitly, following a rescaling of the coordinates $x \rightarrow \lambda x$, $y \rightarrow \lambda y$, $t \rightarrow \lambda^{-1}t$ and $\mu \rightarrow \lambda \mu$. Under this scaling, the energy of the configuration is shifted $E \rightarrow \lambda^4 E$. Setting $\lambda \rightarrow 0$, while at the same time keeping the combination $\lambda \mu$ \textit{finite}, can be achieved by pushing the original mass parameter to large values. Given the above state of-the-art, in the end, we will analyze the complexity in the limit of small $\mu$.

A general state $\ket{\psi (t)}$ can be expanded as
\begin{align}
\label{e4}
    \ket{\psi (t)}=\sum_{n =0}^\infty \frac{(-it)^n}{n !}\ket{\psi_n}.
\end{align}

Here, we introduce a basis $\{ \ket{\psi_n} \}=\{H^n \ket{\psi_0}\} $ that defines the Krylov subspace in the full Hilbert space. In the following, we construct these basis elements by successive application of the Hamiltonian \eqref{e1}.

The initial state $\ket{\psi_0}$ can be expressed as an expansion in the position eigen basis
\begin{align}
\label{e5}
    \ket{\psi_0}=\int dx dy \psi_0 (x,y)\ket{x,y}
\end{align}
such that the basis is orthonormal
\begin{align}
    \braket{x',y'|x,y}=\delta(x-x')\delta (y-y').
\end{align}

Using \eqref{e5}, it is straightforward to show that
\begin{align}
    \braket{\psi_0|\psi_0}=\frac{2 \mu}{\pi}\int dx dy e^{-2\mu (x^2+y^2)}=1.
\end{align}

The first few basis elements can be explicitly constructed through successive applications of the Hamiltonian. They can be expressed as follows.

For $n=1$, we find the following
\begin{align}
    \ket{\psi_1}&=H \ket{\psi_0}\nonumber\\
    &=\int dx dy \phi_1 (x,y)\psi_0(x,y) \ket{x,y}\\
    \phi_1 (x,y)&=2 \mu (1-\mu (x^2 +y^2))+V(x,y).
\end{align}

For $n=2$, we find the following
\begin{align}
    \ket{\psi_2}&=H^2 \ket{\psi_0}\nonumber\\
    &=\int dx dy \phi_2 (x,y)\psi_0(x,y) \ket{x,y}\\
    \phi_2 (x,y)&=\frac{1}{8}F(x,y)+V(x,y)\phi_1(x,y)\\
    F(x,y)&=-32 \left(x^2+y^2\right)+8 \mu  \left(5 \left(x^2+y^2\right)^2+3 y\right)\nonumber\\
    &-\mu ^2 \left(8 \left((x^2+y^2)^3+8 y^3\right)-59\right)\nonumber\\
    &+2 \mu ^3 \left(x^2 \left(8 y^3-61\right)+8 y^5-52 y^2\right)\nonumber\\
    &+6 \mu ^4 \left(x^2+y^2\right) \left(5 x^2+4 y^2\right).
\end{align}

The above procedure can be applied iteratively to construct other states for $n\geq 3$. The next step is to construct an orthonormal Krylov basis $\{ \ket{K_n}\}$ such that $\braket{K_m|K_n}=\delta_{mn}$ and one has an equivalent expansion of the general state \eqref{e4} in terms of Krylov states
\begin{align}
\label{e13}
    \ket{\psi (t)}=\sum_{n =0}^\infty \psi_n(t)\ket{K_n}.
\end{align}

\underline{Gram-Schmidt orthogonalization:}

We propose the following expansion of Krylov basis elements in terms of old basis elements
\begin{align}
    \ket{K_{n+1}}=\ket{\psi_{n+1}}-c_n \ket{K_{n}}-d_{n} \ket{K_{n-1}}
\end{align}
where the coefficients $c_n$ and $d_n$ are to be fixed by orthogonality of the Krylov basis. Here, we choose the initial state as $\ket{K_0}=\ket{\psi_0}$ and list all subsequent states.

For $n=0$, we find the following
\begin{align}
   \ket{K_1}=\ket{\psi_1}-c_0 \ket{K_0}
\end{align}
where we set $\ket{K_{-1}}=0$, without loss of generality.

Next, we take an inner product with $\ket{K_0}$ and set $\braket{K_0|K_1}=0$. This yields the following coefficient
\begin{align}
    c_0=\int dx dy \phi_1(x,y)\psi_0^2 (x,y)=\frac{37 \mu }{32}+\frac{1}{4 \mu ^2}.
\end{align}

Similarly, for $n=1$, we find the following
\begin{align}
   \ket{K_2}=\ket{\psi_2}-c_1 \ket{K_1}-d_1 \ket{K_0}.
\end{align}

Taking the inner product with $\ket{K_0}$, we find
\begin{align}
  \braket{K_0|K_2}=0=\braket{K_0|\psi_2}-  d_1
\end{align}
which yields the following coefficient
\begin{align}
    d_1&=\int dx dy \phi_2(x,y)\psi_0^2(x,y)\nonumber\\
    &=\frac{1}{1024 \mu ^4}(2107 \mu ^6-32 \mu ^3+384).
\end{align}

Similarly, taking the inner product with $\ket{K_1}$ and setting $\braket{K_1|K_1}=N_1$, we find the following condition
\begin{align}
    \braket{K_1|K_2}=0=\braket{K_1|\psi_2}-N_1c_1
\end{align}
which can be further simplified to yield
\begin{align}
   N_1 c_1 &=\int dx dy \phi_1 (x,y)\phi_2(x,y)\psi_0^2(x,y)-c_0 d_1\nonumber\\
    &= \frac{1}{8192 \mu ^6}\Big(27063 \mu ^9-15774 \mu ^6 \nonumber\\
    &+2752 \mu ^3+10752 \Big).
\end{align}

Next, we normalize the Krylov states by rescaling $\ket{K_{n+1}}\rightarrow \frac{1}{\sqrt{N_{n+1}}}\ket{K_{n+1}}$. For $n=0$, we have 
\begin{align}
    \braket{K_1|K_1}&=N_1\nonumber\\
    &=\frac{1}{512 \mu ^4}(369 \mu ^6-312 \mu ^3+160).
\end{align}

On a similar note, for $n=1$, we find the following
\begin{align}
  &\braket{K_2|K_2}=N_2\nonumber\\
    &=\int dx dy \phi^2_2(x,y)\psi_0^2 (x,y)-N_1 c^2_1 -d^2_1.
\end{align}

A straightforward evaluation of the integral reveals
\begin{align}
    &N_2 = \frac{A}{B}\nonumber\\
    &A=280437760 \mu ^6-214646784 \mu ^3+87883776\nonumber\\
    &+3 \mu ^9 \Big(34691409 \mu ^9-38869776 \mu ^6+24120660 \mu ^3\nonumber\\&+14873728\Big)\nonumber\\
    &B=131072 \mu ^8 \left(369 \mu ^6-312 \mu ^3+160\right).
\end{align}

In summary, we have the following set of Krylov states
\begin{align}
\label{e25}
    &\ket{K_0}=\ket{\psi_0}\\
    \label{e26}
    &\ket{K_1}=\frac{1}{\sqrt{N_1}}\Big[\ket{\psi_1} -c_0 \ket{K_0} \Big]\\
    &\ket{K_2}=\frac{1}{\sqrt{N_2}}\Big[\ket{\psi_2}-c_1 \ket{K_1}-d_1 \ket{K_0}\Big].
\end{align}

The Krylov basis $\{ \ket{K_n}\}$ satisfies the Krylov chain condition \cite{Balasubramanian:2022tpr}, which is expressed as
\begin{align}
\label{e28}
    H \ket{K_n}=a_n \ket{K_n}+b_n \ket{K_{n-1}}+b_{n+1}\ket{K_{n+1}}.
\end{align}

Here, $a_n$ and $b_n$ are the Lanczos coefficients, which are given by the following expressions \cite{Hashimoto:2023swv}
\begin{align}
    a_n =\braket{K_n|H|K_n}~;~b_n = \braket{K_{n-1}|H|K_n}.
\end{align}

The coefficients $a_n$ are given by the diagonal element of the Hamiltonian \eqref{e1}. For example, some of these elements can be computed as follows.

Setting, $n=0$, one finds that
\begin{align}
    a_0(\mu)\Big|_{\mu \gg 1}=\braket{K_0|H|K_0}=c_0=1.15625 \mu +\frac{1}{4 \mu ^2}.
\end{align}

Next, we set $n=1$, which yields the following
\begin{align}
  a_1 (\mu)=  \braket{K_1|H|K_1}.
\end{align}

Using \eqref{e26}, one finds that
\begin{align}
\label{e32}
    a_1 (\mu)=\frac{1}{N_1}\Big[\braket{\psi_1|\psi_2}-c_0 \braket{\psi_1|\psi_1}-c_0 d_1+c^3_0  \Big].
\end{align}

A straightforward evaluation of all the above entities in \eqref{e32} yields the following value for the coefficient
\begin{align}
    a_1 (\mu)\Big|_{\mu \gg 1} &= \frac{40473 \mu ^9-22956 \mu ^6+2080 \mu ^3+20224}{11808 \mu ^8-9984 \mu ^5+5120 \mu ^2}\nonumber\\
    &=3.42759 \mu +\mathcal{O}(\mu^{-1}).
\end{align}

Next, we calculate the off-diagonal elements ($b_n$) of the Hamiltonian on the Krylov basis. To begin with, we notice that $b_0=0$. Next, we set $n=1$, which reveals the following Lanczos coefficient
\begin{align}
    b_1(\mu)\Big|_{\mu \gg 1}&=\braket{K_0|H|K_1}\nonumber\\
    &=\frac{1}{16} \sqrt{\frac{80}{\mu ^4}+\frac{369 \mu ^2}{2}-\frac{156}{\mu }}\nonumber\\
    &=0.848942 \mu+\mathcal{O}(\mu^{-1}).
\end{align}

Next, setting $n=2$, one finds the following
\begin{align}
\label{e35}
    &b_2 (\mu)= \braket{K_1 |H|K_2}\nonumber\\
    &=\frac{1}{\sqrt{N_1 N_2}}\Big[ \braket{\psi_1|\psi_3}-d_1 \braket{\psi_1|\psi_1} -c_0 \braket{\psi_0|\psi_3}\nonumber\\
    &+ c^2_0 d_1\Big]-\frac{1}{N_1 \sqrt{N_2}}\Big[c_1\braket{\psi_1|\psi_2} -c_0 c_1 \braket{\psi_1|\psi_1}\nonumber\\
    &-c_0 c_1 d_1 +c^3_0 c_1\Big].
\end{align}

To calculate \eqref{e35}, one has to take note of the state $\ket{\psi_3}$
\begin{align}
\label{e36}
     \ket{\psi_3}&=H^3 \ket{\psi_0}\nonumber\\
    &= \int dx dy \phi_3(x,y)\psi_0 (x,y)\ket{x,y}\nonumber\\
    \phi_3(x,y)&=\frac{1}{32}\mathcal{G}(x,y)+V(x,y)\phi_2 (x,y)
\end{align}
where the detailed expression of $\mathcal{G}(x,y)$ has been provided in the Appendix \ref{ApenA}. 

Using \eqref{e36}, the Lanczos coefficient turns out to be
\begin{align}
    b_2 (\mu) \Big|_{\mu \gg 1}=10.8206 \mu-10.7106+\mathcal{O}(\mu^{-1}).
\end{align}

In an interesting fact, we observe that all non-zero Lanczos coefficients scale linearly with the (mass) deformation parameter in the limit when $\mu \gg 1$, that is, $a_n \sim \mu$ and $b_n \sim \mu$. Later, we complement this observation with its counterpart when the deformation is considered small enough, that is, $\mu \sim 0$.

With the above machinery in hand, the Krylov state complexity is defined as \cite{Balasubramanian:2022tpr}
\begin{align}
    C(t)=\sum_n n | \psi_n(t)|^2.
\end{align}

Using \eqref{e13} and \eqref{e26}, together with 
\begin{align}
    i \partial_t \ket{\psi (t)}=H \ket{\psi (t)}
\end{align}
it is straightforward to show that the wave function $\psi_n(t)$ satisfies the Schrodinger equation \cite{Hashimoto:2023swv}
\begin{align}
\label{e40}
    i \partial_t \psi_n (t)=a_n \psi_n (t)+b_{n+1}\psi_{n+1}(t)+b_n \psi_{n-1}(t).
\end{align}

Setting $n=0$, we obtain the following equation
\begin{align}
\label{e41}
    i \partial_t \psi_0 (t) = a_0 \psi_0(t)+b_1 \psi_1 (t).
\end{align}

To calculate $\psi_1 (t)$, we notice the following expansions that follow receptively from \eqref{e4} and \eqref{e13}
\begin{align}
\label{e42}
    & \ket{\psi(t)}=\ket{\psi_0}-it \ket{\psi_1}+\cdots\\
    & \ket{\psi(t)}=\psi_0(t)\ket{K_0}+\psi_1(t)\ket{K_1}+\cdots 
    \label{e43}
\end{align}
where the Krylov states are defined in \eqref{e25} and \eqref{e26}.

Using \eqref{e42}-\eqref{e43} and thereby taking the inner product with $\ket{K_1}$, we finally obtain
\begin{align}
    \psi_1(t)=-\frac{it}{\sqrt{N_1}}\Big[ \int dx dy \phi_1^2(x,y)\psi_0^2(x,y)-c^2_0 \Big].
\end{align}

After performing the integral, we finally obtain
\begin{align}
\label{e45}
    \psi_1(t)=-i  b_1 t.
\end{align}

Substituting \eqref{e45} into \eqref{e41}, we find
\begin{align}
    \psi_0 (t)=\frac{b_1^2 }{a^2_0}(-1+i a_0 t)+ e^{-i a_0 t}
\end{align}
where we set the constant of integration to one.

Next, we set $n=1$ in \eqref{e40}, which yields the following equation for the wave function $\psi_n(t)$
\begin{align}
    i \partial_t \psi_1 (t)=a_1 \psi_1(t)+b_2 \psi_2(t)+b_1 \psi_0(t).
\end{align}

After simplification, one finds the following
\begin{align}
\label{e48}
    \psi_2(t)=\frac{b_1}{b_2}-\frac{a_1}{b_2}\psi_1(t)-\frac{b_1}{b_2}\psi_0(t).
\end{align}

Using \eqref{e45} and \eqref{e48}, the first two terms in the Krylov complexity yields the following
\begin{align}
\label{e49}
    C(t)|_{t \sim 0}&=| \psi_1(t)|^2+2| \psi_2(t)|^2 + \cdots \nonumber\\
    &=C_0 +C_1 t^2 +C_2(t)+\cdots.
\end{align}

Individual entities can be expressed as
\begin{align}
    &C_0 = \frac{2 b_1^6}{a_0^4 b_2^2}+\frac{4 b_1^4}{a_0^2 b_2^2}+\frac{4 b_1^2}{b_2^2}\\
    &C_1=b^2_1\Big[1+\frac{2  \left(b_1^2-a_0 a_1\right)^2}{a_0^2 b_2^2}\Big]=\beta \mu^2\\
    &C_2(t)=\frac{4 b_1^2 \left(a_0^2+b_1^2\right) }{a_0^2 b_2^2}\cos (a_0 t)\nonumber\\
    &+\frac{4 b_1^2  \left(a_0 a_1-b_1^2\right) }{a_0 b_2^2}t\sin (a_0 t).
\end{align}

\begin{figure}
    \centering
    \includegraphics[width=0.7\linewidth]{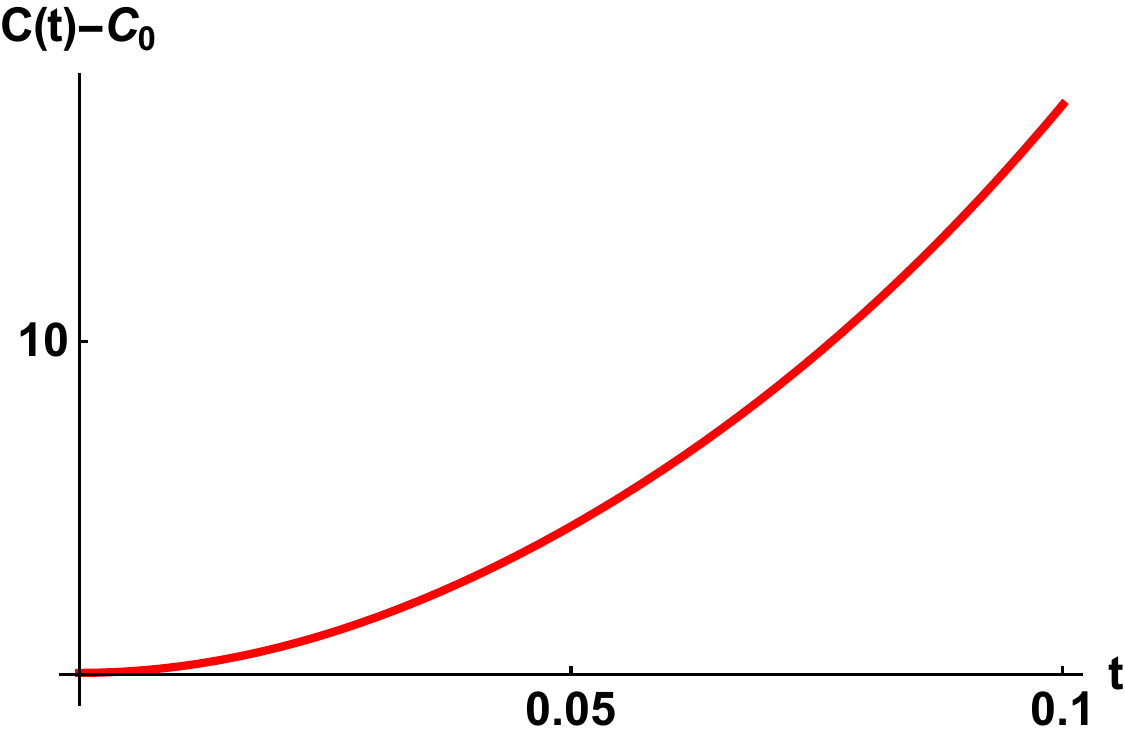}
    \caption{We plot complexity in the large deformation limit, where we set $\mu=50$, $\beta=0.7$, $\gamma = 0.05$ and $\zeta=0.08$.}
    \label{figcmu}
\end{figure}

Clearly, in the large deformation ($\mu \gg 1$) limit, where the Lanczos coefficients scale linearly with the deformation parameter, that is, $a_n,b_n \sim \mu$, the leading coefficient $C_0$ is independent of $\mu$. On the other hand, the sub-leading term scales quadratically, that is, $C_1 \sim \mu^2$. Following similar arguments, the third coefficient in the expansion \eqref{e49} can be expressed following some suitable rescaling of the parameters as
\begin{align}
    C_2(t)=\gamma \cos(\mu t)+\zeta \mu t \sin(\mu t).
\end{align}

In Fig.\ref{figcmu} we plot the first few terms in \eqref{e49}. As the plot reveals, the complexity growth at initial time is quadratic in nature. Similar observation has been made from holography in \cite{Roychowdhury:2026sgg}, in the context of operator complexity.
\section{Small deformation limit}
\label{sec3}
We now move towards the other corner of the parameter space where we treat the mass parameter $\mu \sim 0$ as a perturbation to the BFSS matrix model \cite{Banks:1996vh}. To begin with, we consider the case with $\mu =0$, which corresponds to an integrable sector \cite{Asano:2015eha},\cite{Amore:2024ihm} of the full theory. 

The corresponding Hamiltonian is given by
\begin{align}
\label{e55}
    H_{BFSS}(x,y)=\frac{p^2_x}{2}+\frac{p^2_y}{2}+\frac{1}{2}(x^2+y^2)^2.
\end{align}

To obtain the ground state wave function (which is our starting ansatz for $n=0$), one has to solve the time independent Schrodinger equation 
\begin{align}
\label{e56}
    \Big[\partial^2_x+\partial^2_y-(x^2+y^2)^2\Big]\Psi_n(x,y)=-2E_n \Psi_n(x,y).
\end{align}

We solve \eqref{e56} in polar coordinates ($r, \theta$), where we propose an ansatz of the following form
\begin{align}
    \Psi_{n,\ell} (r, \theta)=e^{i \ell  \theta}\psi_n (r).
\end{align}

The radial wave function $\psi_n(r)$ satisfies an equation of the following form
\begin{align}
    \partial^2_r \psi_n(r)+\frac{1}{r}\partial_r \psi_n(r)+\Big[2E_n -V_{eff} (r)\Big]\psi_n(r)=0.
\end{align}

The effective potential can be expressed as
\begin{align}
\label{e59}
    V_{eff}(r)=\frac{\ell^2}{r^2}+r^4
\end{align}
where $\ell$ is the angular momentum quantum number. 

Notice that the potential \eqref{e59} has a minimum at $r_0\sim \ell^{1/3}$. We look for a configuration close to the minima of the potential. Following a rescaling of the coordinates as $r \rightarrow \lambda r$, $t \rightarrow \lambda^{-1}t$, $\ell \rightarrow \lambda^3 \ell$, reveals an overall scaling of the total energy of the configuration as
\begin{align}
    E=\frac{\dot{r}^2}{2}+\frac{\ell^2}{2 r^2}+\frac{r^4}{2} \rightarrow \lambda^4 E.
\end{align}

Clearly, the energy minima of the configuration can be obtained in the limit $\lambda \rightarrow 0$, while at the same time setting $\ell = 0$. This would result in a free particle wave function in polar coordinates 
\begin{align}
    \psi_0(r)=N J_0(r)
\end{align}
where we set $E_0=\frac{1}{2}$ for simplicity. 

Notice that the wave function vanishes at large distances, that is, $\psi_0(r)|_{r\rightarrow \infty}=0$. The state corresponding to $n=0$ (and therefore at $t=0$) is defined as
\begin{align}
    &\ket{\Psi_0}=\int d^2x  \Psi_{0}(r)\ket{r,\theta}=N \int d^2 x J_0(r)\ket{r,\theta}\\
     &\braket{r', \theta'|r,\theta}=\frac{1}{r'}\delta (r-r')\delta (\theta - \theta').
\end{align}

The constant $N=\frac{1}{\sqrt{\pi}R J_1(R)}$ is fixed by normalization and the boundary condition $J_0(R)=0$, which yields
\begin{align}
\label{e62}
   \int_0^{2\pi} d\theta \int_0^R dr~ r| \Psi_{0}(r)|^2=1
\end{align}
where the particle is confined within a disk of radius $R$.

Finally, the $n=0$ state can be expressed as
\begin{align}
\label{e65}
    \ket{\Psi_{0}}=\frac{1}{\sqrt{\pi}R J_1(R)}\int d^2 x J_0(r)\ket{r,\theta}
\end{align}
which clearly satisfies $\braket{\Psi_0 |\Psi_0}=1$ by virtue of \eqref{e62}. 

In what follows, we explore the time evolution of the state \eqref{e65} under \eqref{e1}. The subsequent state corresponding to $n=m$ is obtained by applying the Hamiltonian
\begin{align}
    \ket{\Psi_m}&=H^m \ket{\Psi_0}\\
    H&=-\frac{1}{2}\Big(\partial^2_r +\frac{1}{r}\partial_r+\frac{1}{r^2}\partial^2_\theta \Big)+\frac{r^4}{2}\nonumber\\
    &+\frac{\mu^2 r^2}{8}(1+3 \sin^2\theta)-\mu r^3 \sin^3 \theta\nonumber\\
    &=-\frac{1}{2}\Big(\partial^2_r +\frac{1}{r}\partial_r+\frac{1}{r^2}\partial^2_\theta \Big)+V(r,\theta)
\end{align}

Notice that under the scaling $r \rightarrow \lambda r$ and $\mu \rightarrow \lambda \mu$, the potential term scales to $V(r,\theta)\sim \lambda^4 V(r,\theta)$, where $\lambda \rightarrow 0$. In the following, we construct states that receive corrections up to $\mathcal{O}(\lambda^4)$, under the above scaling.

A straightforward computation reveals (for $m=1,2$)
\begin{align}
    &\ket{\Psi_1}=N \int d^2x f(r) J_{0}(r)\ket{r,\theta}\nonumber\\
    &f(r)=\frac{1}{8}\Big[4+4 r^4+\mu  r^2 \sin ^2\theta  (3 \mu -8 r \sin \theta )+\mu ^2 r^2\Big]\nonumber\\
    &=\frac{1}{2}+V(r,\theta)\\
    &\ket{\Psi_2}=N\int d^2x \Psi_2(r)\ket{r,\theta}\\
    &\Psi_2(r)=\frac{1}{4} \left(r^4-16 r^2+1\right) J_0(r)+\frac{1}{2}V(r,\theta)J_0(r)\nonumber\\
    &+2 r^3 J_1(r)+\frac{\mu}{2} r \sin \theta \Big[ 6 J_0(r)-r\sin^2\theta \Big( r J_0(r)\nonumber\\
    &+6 J_1(r)\Big)\Big]+\frac{\mu^2}{32}\Big[ 4 r (5-3 \cos (2 \theta )) J_1(r)\nonumber\\
    &+J_0(r) \left(5 \left(r^2-4\right)-3 r^2 \cos (2 \theta )\right)\Big].
\end{align}

We write down (normalized) Krylov states as follows
\begin{align}
    & \ket{K_0}=\ket{\Psi_0}\\
    \label{e72}
    &\ket{K_1}=\frac{1}{\sqrt{N_1}}\Big[\ket{\Psi_1}-c_0 \ket{K_0} \Big]\\
    \label{e73}
    &\ket{K_2}=\frac{1}{\sqrt{N_2}}\Big[\ket{\Psi_2}-c_1 \ket{K_1}-d_1 \ket{K_0}\Big].
\end{align}

The constants $c_i$ and $d_i$ are obtained using normalization of the Krylov states $\braket{K_m|K_n}=\delta_{mn}$, which yields
\begin{align}
    &c_0=\braket{\Psi_0|\Psi_1}=\frac{1}{96 J^2_1(R)}\Big[48 J_1(R){}^2 +16 R^4\nonumber\\
    & \times \, _2F_3\left(\frac{1}{2},3;1,1,4;-R^2\right)+15 \mu ^2 R^2 \nonumber\\
    & \times \, _2F_3\left(\frac{1}{2},2;1,1,3;-R^2\right)\Big]\\
\label{e75}
&c_1 N_1=\braket{\Psi_1|\Psi_2}-c_0 \braket{\Psi_0|\Psi_2}\nonumber\\
&=\frac{1}{8}+\mathcal{F}(R)-c_0 d_1\\
&d_1=\frac{1}{240} \Big[4 \left(6 R^4-32 R^2+79\right)\nonumber\\
&+25 \mu ^2 \left(R^2-2\right)\Big]
\label{e76}.
\end{align}
Detailed expressions pertaining to \eqref{e75} along with normalization constants are provided in Appendix \ref{ApenA}.

\underline{Lanczos coefficients:} In the following, we note the first few Lanczos coefficients. The Lanczos coefficients that sit at the diagonal are given by
\begin{align}
    &a_0 (\mu ,R)=c_0\\
    &a_1 (\mu , R)=\frac{1}{N_1}\Big[\braket{\Psi_1|\Psi_2}-c_0 \braket{\Psi_1|\Psi_1}\nonumber\\
    &-c_0 d_1+c^3_0  \Big].
\end{align}

\begin{figure}
    \centering
    \includegraphics[width=1\linewidth]{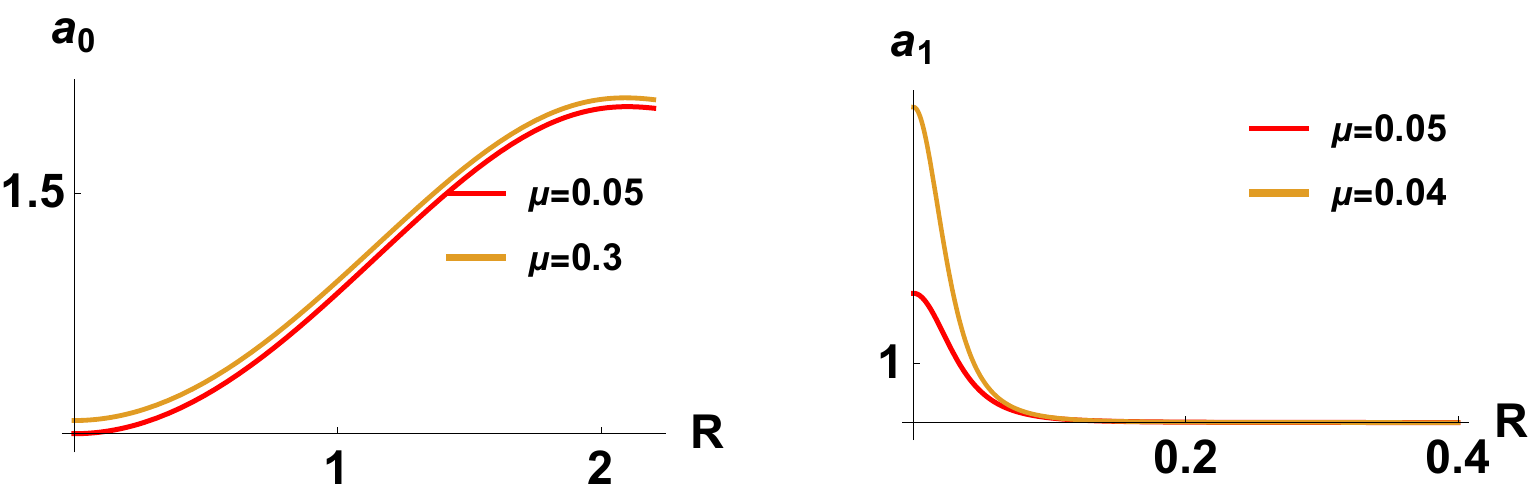}
    \caption{We plot the Lanczos coefficients $a_0$ and $a_1$ as a function of the size $R$ of the disk for different $\mu$.}
    \label{figaR}
\end{figure}

\begin{figure}
    \centering
    \includegraphics[width=0.7\linewidth]{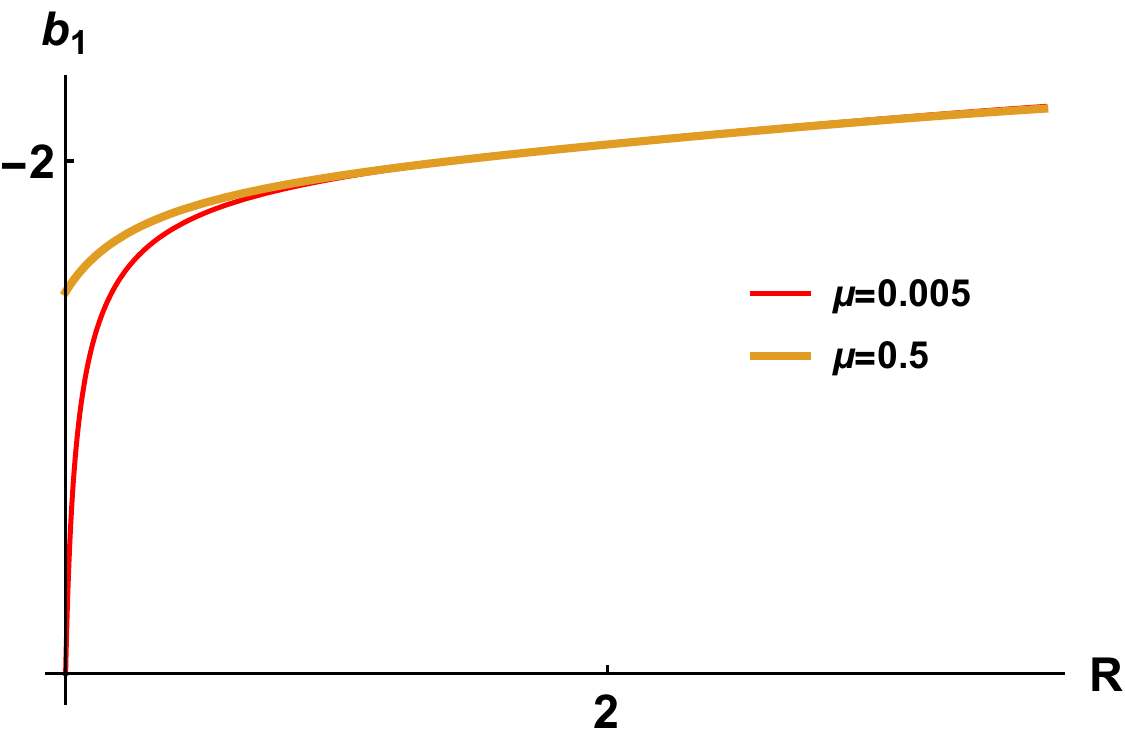}
    \caption{We plot the Lanczos coefficient $b_1$ as a function of the size $R$ of the system for different $\mu$.}
    \label{figbR}
\end{figure}

A careful analysis reveals that, unlike the strong deformation limit, the LO correction to diagonal Lanczos coefficients appears in quadratic order in the mass deformation, that is, $a_0,a_1 \sim \mu^2$. In Fig.\ref{figaR} we plot $a_0$ for different choices of the mass parameter $\mu$, which shows that it increases with $\mu$. In addition, it reaches saturation as the size of the disk grows large enough. 

Reverse behavior is observed for the coefficient $a_1$, which, instead of increasing with size, starts to fall sharply for larger $R$, thus achieving saturation ($\sim 0.05$) for large size of the disk. A closer analysis further reveals that its value is lower for a larger mass deformation $\mu$ as opposed to the coefficient $a_0$. 

On the other hand, off diagonal elements are given by
\begin{align}
    &b_1 (\mu , R)=\frac{1}{\sqrt{N_1}}\Big[d_1-c^2_0 \Big].
\end{align}

Like diagonal elements $a_0 ,a_1$, the off-diagonal Lanczos coefficient $b_1$ is also modified at LO in $\mu^2$. As Fig.\ref{figbR} reveals, the coefficient $b_1$ has a sharp increase and saturates rapidly with increasing size of the disk. In Fig.\ref{figbR}, we plot the coefficient $b_1$ for different choices of the parameter, which clearly reveals universality, in the sense that for large size of the disk they saturate to a constant $\sim -2$.

Notice that in the original formulation of the BMN matrix model \cite{Asano:2015eha}, one should think of the matrices expressed in $2 \times 2$ representation as $X^i=r(t)\frac{\sigma^i}{2}$, where $\sigma^i$ are the Pauli matrices. Clearly, large size ($R \rightarrow \infty$) of the disk would correspond to matrices with large entries, which saturates the corresponding Lanzcos coefficients $a_n ,b_n$. Ideally, this is the regime where the complexity in the small deformation limit should be explored. Like before, the leading term in complexity is quadratic in $t$ and is controlled by the parameter $b_1$, that is, $C(t) \sim b^2_1 t^2$.
\section{Conclusion and outlook}
\label{sec4}
In summary, we compute the Lanczos coefficients in the BMN matrix model considering a particular reduction ansatz, known as the pulsating fuzzy sphere model \cite{Asano:2015eha}. We develop an analytic technique to compute the first few Lanczos coefficients, and hence the early growth in time of complexity in both large and small mass deformations. In the large deformation limit, the initial (or minimum energy) configuration corresponds to two localized harmonic oscillators \cite{Amore:2024ihm} that evolve under the BMN Hamiltonian. On the other hand, in the small deformation limit, the minimum energy configuration is identified as the free particle Hamiltonian that is (perturbatively) corrected due to massive deformation. 

In the limit of large mass deformation $\mu \gg 1$, the Lanczos coefficients appear to be growing linearly with the deformation parameter, that is, $a_n, b_n \sim \mu$. On the other hand, in the small deformation limit, the Lanczos coefficients quickly saturate as the size of the disk becomes large, and the first nontrivial correction appears at order $\mu^2$ where $\mu \ll 1$. The above behavior appears to be a universal characteristic for a wide range of values of $\mu$.

It would be nice to extend the above analysis for intermediate coupling $\mu$, and, in particular, to examine the scaling of Lanczos coefficients with mass parameter $\mu$. As an immediate extension, one could look for other systematic reductions, that is, $N=3$ and $N=4$ matrix models \cite{Asano:2015eha}, where a similar analysis can be performed. Finally, an insight from holography will complete the cycle. In particular, it would be nice to identify a gravitational counterpart of the calculations presented in this paper and to find a systematic way to extract Lanczos coefficients for an intermediate coupling of the matrix model. We hope to address these issues in the near future.
\section*{Acknowledgments} The author thanks Carlos Nunez for his comments on the draft. The author also acknowledges the Mathematical Research Impact Centric Support (MATRICS) grant (MTR/2023/000005) received from ANRF, India.
\appendix
\section{Detailed expressions}
\label{ApenA}
The detailed expression of $\mathcal{G}(x,y)$ is provided below.
\begin{align}
  \mathcal{G}(x,y)&=-256 \left(\left(x^2+y^2\right)^3-1\right)+16 \mu  \left(x^2+y^2\right) \Big(9 x^6\nonumber\\
    &+27 x^4 y^2+3 x^2 y \left(9 y^3+2\right)+9 y^6+46 y^3-144\Big)\nonumber\\
    &-4 \mu^2 \Big( 4 x^{10}+20 x^8 y^2-643 y^4-336 y\nonumber\\
    &+40 x^6 y^4+4 \left(y^3+32\right) y^7+2 x^2 y^2 \Big(10 y^6\nonumber\\
    &+128 y^3-787\Big)+x^4 \left(40 y^6+128 y^3-811\right)\Big)\nonumber\\
    &+8 \mu^3 \Big( 8 y^9-76 y^6-479 y^3+167+6 x^4 y^2\nonumber\\
    &(4 y^3-73)+x^6 \left(8 y^3-153\right)+3 x^2 y\nonumber\\
    &(8 y^6-139 y^3-23)\Big)+\mu^4 \Big(120 x^8+456 x^6 y^2\nonumber\\
    &+648 x^4 y^4 +x^2 \left(344 y^6+2080 y^3-4207\right)\nonumber\\
    &+4 y^2 \left(8 y^3 \left(y^3+56\right)-847\right)\Big)\nonumber\\
    &-3\mu^5 \left(5 x^2+4 y^2\right)\Big(4 y^2 \left(4 y^3-31\right)+x^2\nonumber\\
    &(16 y^3-139)\Big)-9 \mu^6 (x^2+y^2)\left(5 x^2+4 y^2\right)^2.  
\end{align}

Detailed expression of $\mathcal{F}(R)$ \eqref{e75} is given below.
\begin{align}
    & \mathcal{F}(R)=\frac{1}{40J_1(R){}^2}\Big[ 5 R^4 \, _2F_3\left(\frac{1}{2},3;1,1,4;-R^2\right)\nonumber\\
    &+20 R^6 \, _2F_3\left(\frac{1}{2},4;1,1,5;-R^2\right)+2 R^8 \nonumber\\
    &\times \, _2F_3\left(\frac{1}{2},5;1,1,6;-R^2\right)\Big]+\frac{\mu^2}{384 J_1(R){}^2}\nonumber\\
    &\times \Big[45 R^2 \, _2F_3\left(\frac{1}{2},2;1,1,3;-R^2\right)+376 R^4 \nonumber\\
    & \times \, _2F_3\left(\frac{1}{2},3;1,1,4;-R^2\right)+60 R^6\nonumber\\
    &\times \, _2F_3\left(\frac{1}{2},4;1,1,5;-R^2\right) \Big]+\mathcal{O}(\mu^4).
\end{align}

Normalization constants in \eqref{e72} and \eqref{e73} are given by
\begin{align}
    &N_1=\braket{\Psi_1|\Psi_1}-c^2_0\nonumber\\
    &=\frac{1}{4}+\frac{1}{60 J_1(R){}^2}\Big[10 R^4 \, _2F_3\left(\frac{1}{2},3;1,1,4;-R^2\right)\nonumber\\
    &+3 R^8 \, _2F_3\left(\frac{1}{2},5;1,1,6;-R^2\right)\Big]+\frac{5\mu^2 R^2}{32 J_1(R){}^2}\nonumber\\
    & \times \Big[ \, _2F_3\left(\frac{1}{2},2;1,1,3;-R^2\right)+R^4 \nonumber\\
    & \times \, _2F_3\left(\frac{1}{2},4;1,1,5;-R^2\right)\Big]-c^2_0+\mathcal{O}(\mu^4).\\
    &N_2=\braket{\Psi_2|\Psi_2}-c^2_1 N_1-d^2_1\nonumber\\
    &=\frac{1}{5040}\Big[ 4 R^2 \left(35 R^6+800 R^4+1503 R^2-8016\right) \nonumber\\
    &+64443\Big]+\frac{\mu^2}{3360}\Big[300 R^6+4278 R^4\nonumber\\
    &+3679 R^2-7358 \Big]-c^2_1 N_1-d^2_1+\mathcal{O}(\mu^4).
\end{align}


\end{document}